\documentclass[a4paper,11pt]{article}
\usepackage{latexsym}
\usepackage{amsmath, amsthm, amssymb, bbm}
\usepackage[mathscr]{eucal}
\usepackage{hyperref}
\usepackage{slashed}
\usepackage{graphicx}

\theoremstyle{plain}

\theoremstyle{definition}

\theoremstyle{remark}

\newcommand{\rhs}{r.h.s.\ }

\newcommand{\wrt}{w.r.t.\ }
\newcommand{\cf}{cf.\ }

\newcommand{\ud}{\mathrm{d}}
\newcommand{\del}{\partial}

\newcommand{\betrag}[1]{{\lvert #1 \rvert}}

\newcommand{\order}{\mathcal{O}}

\newcommand{\eps}{\varepsilon}

\DeclareMathOperator{\Ei}{Ei}

\DeclareMathOperator{\tr}{tr}

\newcommand{\s}[1]{{\underline{#1}}}

\newcommand{\nabslash}{{\slashed \nabla}}

\begin{document}

\title{The current density in quantum electrodynamics in time-dependent external potentials and the Schwinger effect}
\author{Jochen Zahn \\ Institut f\"ur Theoretische Physik, Universit\"at Leipzig \\ Br\"uderstr.~16, 04103 Leipzig, Germany \\ jochen.zahn@itp.uni-leipzig.de}

\date{\today}

\maketitle

\begin{abstract}
In the framework of quantum electrodynamics (QED) in external potentials, we introduce a method to compute the time-dependence of the expectation value of the current density for time-dependent homogeneous external electric fields. We apply it to the so-called Sauter pulse. For late times, our results agree with the asymptotic value due to electron-positron pair production. We correct, and compare to, a general expression derived by Serber for the linearization in the external field.
Based on the properties of the current density, we argue that the appearance of enhanced quasi-particle densities at intermediate times in slowly varying sub-critical potentials is generic. Also an alternative approach, which circumvents these difficulties, is sketched.
\end{abstract}
{\bf Classification (PACS):} 12.20.Ds; 11.40.-q 

\section{Introduction}

A crucial prediction of QED in external potentials is the production of electron-positron pairs in sufficiently strong electric fields \cite{Sauter31, EulerHeisenberg, Schwinger51}, the so-called \emph{Schwinger effect}. This is a non-perturbative effect which is exponentially suppressed for field strengths below the critical field strength $E_c = m^2/\betrag{e} \sim 10^{16} \mathrm{V}/\mathrm{cm}$. Hence, it seems unrealistic to test it with static fields. More promising are high intensity lasers, \cf the reviews \cite{Dunne08, RuffiniVereshchaginXue10} and references therein. However, even the projected high intensity laser facilities may not reach this non-perturbative regime. One possibility to overcome this is to use complicated field configurations, such as in the dynamically assisted Schwinger effect \cite{SchutzholdGiesDunne08}. Often, such configurations can only be treated numerically. One framework for this is the quantum kinetic equation (QKE) \cite{QuantumKineticEquation98}, where one defines quasi-particle number densities at intermediate times and obtains the asymptotic particle number by solving an ordinary differential equation for their time evolution.\footnote{To be precise, in order to avoid an integro-differential equation, one either has to supplement the quasi-particle number density with other variables, as in \eqref{eq:n_eom} below, or use the underlying Bogoliubov coefficients, \cf \cite{HebenstreitThesis} for a detailed discussion.} It was noticed \cite{BlaschkeEtAl06}, that the quasi-particle numbers, as defined in this framework, can be much larger at intermediate times than at asymptotically late times (for parameters typical for an optical laser setting, the proportionality was $10^{11}$). This is problematic in numerical studies, as the errors that are made at intermediate times, where the number densities are high, are propagated to asymptotic times, where they can dominate the actual number densities.

Now the concept of a particle is ambiguous at intermediate times, where the external potential is non-vanishing and varying; consequently, one speaks of \emph{quasi-particles}.\footnote{For a recent discussion of these ambiguities in the context of scalar QED in external potentials, we refer to \cite{DabrowskiDunne14}. In the context of quantum field theory on curved space-times, these issues were discussed for example in \cite{Parker69}, \cite{Fulling79}.} One may thus wonder how generic the occurrence of enhanced quasi-particle densities at intermediate times is. Or more generally: consider any formalism in which particle creation is studied by (numerically) solving an evolution equation for some dynamical quantity.\footnote{Another such framework is the Wigner function formalism introduced in \cite{BialynickiBirulaEtAl91}, \cf \cite{HebenstreitEtAl11}, for example. Compared to the quantum kinetic equation it has the advantage of being in principle applicable also when the external field is more general than a homogeneous electric field.} Can one make any generic statements about the intermediate time behavior? One possibility to investigate this is to consider a quantity that is unique and well-defined also at intermediate times. The prime candidate for this is the current density\footnote{The analog of this for quantum field theory in curved space-times would be the stress-energy tensor.}
\begin{equation}
\label{eq:CurrentDensity}
 j^\mu = e \bar \psi \gamma^\mu \psi.
\end{equation}
It is an observable, and thus, at least in principle, measurable. Such a measurement would proceed via back-reaction, i.e., by the changes in the electromagnetic field which are sourced by this current. These back-reaction effects can be used to constrain the generation of primordial electromagnetic fields during inflation \cite{KobayashiAfshordi14}. However, in the present work we are more interested in generic features of this observable and whether these may help to better understand the intermediate time behavior of the dynamical variable in the time evolution equation in which the system is studied. In particular, we will argue that the occurrence of enhanced quasi-particle numbers at intermediate times is generic, and will also indicate a new scheme which avoids these difficulties.

The current density \eqref{eq:CurrentDensity} is unique in the following sense: For a locally gauge covariant definition of the current density, the only ambiguity consists in adding multiples of the external current density \cite{LocCovDirac}, corresponding to a charge renormalization. It can be fixed by requiring that for static external potentials the total vacuum polarization charge vanishes at linear order in the external potential \cite{Heisenberg34, EulerHeisenberg}, i.e., the external charges are not renormalized at this order of approximation.

The expectation value of the current density \eqref{eq:CurrentDensity}, henceforth also called the \emph{vacuum current density}, is a crucial ingredient in the study of back-reaction effects. In the QKE framework, an expression for the vacuum current density was given in \cite{BlochEtAl99}, \cf also \cite{KlugerEtAl92}. In fact, the analysis of this expression will be crucial for the explanation of the enhanced quasi-particle numbers in slowly varying sub-critical potentials reported in \cite{BlaschkeEtAl06}. However, the renormalization prescription used there is not explicitly local, so it is a priori not clear whether it conforms to the requirement of local gauge covariance.\footnote{We refer to \cite{Current} for an elementary example showing that in the presence of external potentials a seemingly sensible but not explicitly local renormalization prescription can give wrong results.} It it thus desirable to check these results with independent methods.

Based on Dirac's method of point-splitting \wrt the singular part of the two-point function,\footnote{This is the analog of the covariant point-splitting renormalization of the stress-energy tensor in the context of quantum field theory in curved space-times \cite{Christensen76}, \cite{WaldTraceAnomaly}, \cf also \cite{HollandsWaldReview} for a recent review.} we give, in Section~\ref{sec:Current}, a general expression for the vacuum current in a homogeneous time-dependent external electric field. We show that it is equivalent to the expression derived in the QKE framework. We also evaluate it for the concrete case of the so-called Sauter pulse,
\begin{align}
\label{eq:Potential}
 E(t) & = \frac{E_0}{\cosh^2 w t}, &
 A(t) & = - \frac{E_0}{w} \tanh w t,
\end{align}
both for super- and sub-critical peak field strengths $E_0$.
This external field was studied extensively in the literature, with an emphasis on the asymptotic behavior, i.e., the pair production probability, which was first computed by Narozhnyi and Nikishov \cite{NarozhnyiNikishov70}. The intermediate time behavior of the quasi-particle number in this system was studied in \cite{HebenstreitThesis, BlaschkeEtAl13},\footnote{In the former reference, no physical interpretation of the quasi-particle number at intermediate times was attempted, noting that an interpretation as real particles is only possible at asymptotic times.} using the quantum kinetic equation \cite{QuantumKineticEquation98}. Our results for the Sauter pulse agree well with the asymptotic pair production probabilities derived in \cite{NarozhnyiNikishov70}. 

We also recall, in Section~\ref{sec:Linearization}, that there is a general expression for the vacuum current density at linear order in the external potential, due to Serber \cite{Serber35}. However, his derivation contains some mistakes, leading to an incorrect result, which we correct here. For sub-critical peak field strengths, the application of this general expression to the Sauter pulse \eqref{eq:Potential} leads to a nice agreement with the results obtained by our new method. One interesting feature of Serber's general expression is that, at the linear order, the vacuum current only depends on the external current, not on the field strength.

We remark that the study of generic properties of the current density does not seem to have attracted a lot of attention in the literature. Usually, the current density is considered in the context of back-reaction, with an emphasis on directly including these effects in the quantum evolution equation \cite{KlugerEtAl92, KlugerMottolaEisenberg98, BlochEtAl99}. For a work where the vacuum current was computed independently of back-reaction, we refer to \cite{GavrilovGitman08}.

Finally, in Section~\ref{sec:Discussion}, we turn to the discussion of the enhanced quasi-particle numbers reported in \cite{BlaschkeEtAl06} for  slowly varying sub-critical potentials. For the adiabatic basis usually considered, this turns out to be a straightforward consequence of the expression for the vacuum current density in terms of the quasi-particle density. We argue that this is indeed a generic feature. Also a new framework, based on Dirac's point splitting method, is sketched, in which these problems should be absent, and which is in principle also applicable for generic external fields.

\section{The current density}
\label{sec:Current}

For the computation of the vacuum current density we use Dirac's method \cite{Dirac34} of point-splitting \wrt the singular part of the two-point function $\langle \psi(x) \bar \psi(y) \rangle$, the so-called \emph{Hadamard parametrix} $H(x,y)$, \cf \cite{Current} for a detailed discussion. This method is conceptually appealing as it is explicitly local and gauge covariant. The practical difficulty is that the Hadamard parametrix is known in position space, while the two-point function is typically given by an integral over modes. The subtraction of the singular part $H$ from this mode integral is in general quite challenging. In \cite{Current} a solution to this problem for the computation of the vacuum polarization in static external potentials was proposed. The idea is to consider the limit of coinciding points from the time direction and to rewrite $\tr \gamma^0 H(x, x + s e^0)$ (up to a potential logarithmic singularity in $s$) in the form of an integral over (shifted) vacuum mode densities. The subtraction of these singular parts can thus be performed inside the mode integral, which renders it much less divergent and numerically more tractable.

This method can be adapted to the case of a homogeneous external electric field along the $z$-axis, in the gauge $A^\mu(x) = \delta_3^\mu A(x^0)$. It is then advantageous to perform the limit of coinciding points from the $z$-direction. An analogous rewriting of the Hadamard parametrix yields the following expression for the vacuum current:
\begin{multline}
\label{eq:j}
 \langle j^3(x) \rangle = e \lim_{z\to 0} \left[ \int \ud p_3 \ e^{i p_3 z} \left( f(p_3,x^0) - f_0(p_3 - e A(x^0),x^0) \right) \right. \\ \left. + \vphantom{\int} \tfrac{1}{12 \pi^2} e J^{3}(x^0) \left( \log z^2/\Lambda^2 - 1 \right) \right].
\end{multline}
Here $f$ is defined as follows. Consider the properly normalized mode solutions $\psi^{\pm}_{\s{p}, s}(x) = f^\pm_{\s{p}, s}(x^0) e^{i \s{p} \s{x}}$ of the Dirac equation
\[
 (i \nabslash - m) \psi = 0
\]
in the presence of the external potential, which coincide with the positive/negative energy modes at asymptotically early times (we refer to Appendix~\ref{app:RelationQKEPointSplit} for a precise definition of the required normalization). Here $s$ is a spin label, $\s{p}$ a three-momentum, and $x = (x^0, \s{x})$. Then $f(p_3, x^0)$ is defined as
\begin{equation*}
 f(p_3, x^0) = \sum_s \int \ud^2 p_\perp \ \overline{f^+_{\s{p}, s}(x^0)} \gamma^3 f^+_{\s{p}, s}(x^0),
\end{equation*}
where $p_\perp = (p_1, p_2)$ is the perpendicular momentum. The function $f_0$ in \eqref{eq:j} is defined analogously, but for the standard mode solutions $f^{0,+}_{\s{p}, s}$ in the absence of external potentials. The corresponding term in \eqref{eq:j} can thus be seen as the subtraction of the vacuum contribution, with an appropriate shift in the momentum.\footnote{One may wonder whether there is a gauge dependence in \eqref{eq:j} due to the occurrence of $A(x^0)$. However, a shift $A \to A + a$, has to be accompanied by the replacement $f(p_3,x^0) \to f(p_3 - e a, x^0)$ in \eqref{eq:j}, as the new solution to the equation of motion \eqref{eq:f_eom} for the mode functions is given by $f'^\pm_{\s{p},s} = f^\pm_{\s{p} - e a e_3, s}$. With the substitution $q_3 = p_3 - e a$, this results in a phase shift $e^{i e a z}$ of the integral in \eqref{eq:j}, which is irrelevant in the limit $z \to 0$.} The third term in \eqref{eq:j} cancels a remaining logarithmic divergence in the presence of an external current $J^3 = \del_\lambda F^{3 \lambda}$. As discussed in the introduction, we fix the length scale $\Lambda$ such that external charges are not renormalized at the linear level, i.e.,
\begin{equation}
\label{eq:Lambda}
\Lambda=2/(e^\gamma m),
\end{equation}
\cf \cite{EulerHeisenberg}, with $\gamma$ the Euler-Mascheroni constant. In fact, using the relation \eqref{eq:CountertermRelation}, \eqref{eq:j} may be rewritten as
\[
  \langle j^3(x) \rangle = e \int \ud^3 \s{p} \left( \sum_s \overline{f^+_{\s{p},s}} \gamma^3 f^+_{\s{p},s} - \sum_s \overline{f^{0,+}_{\s{\pi},s}} \gamma^3 f^{0,+}_{\s{\pi},s} - \frac{e J^3}{(2 \pi)^3} \frac{\eps_\perp^2}{2 \omega_\s{\pi}^5} \right),
\]
where
\begin{align}
\label{eq:def_pi}
 \pi_{1/2} & = p_{1/2}, \\
 \pi_3 & = p_3 - e A(x^0), \nonumber \\
 \eps_\perp^2 & = m^2 + p_1^2 + p_2^2, \nonumber \\
 \omega_\s{p}^2 & = m^2 + \betrag{\s{p}}^2. \nonumber
\end{align}

There is also an expression for the vacuum current density in the framework of the QKE,
\begin{equation}
\label{eq:QKE_current}
 \langle j^3(t) \rangle = - 2 e \int \frac{\ud^3 \s{p}}{(2 \pi)^3} \left[ \frac{\pi_3}{\omega_\s{\pi}} n(\s{p}) + \frac{\omega_\s{\pi}}{e E(t)} \dot n(\s{p}) - \frac{e \dot E(t) \eps_\perp^2}{4 \omega_\s{\pi}^5} \right]
\end{equation}
due to \cite{BlochEtAl99}, \cf also \cite{HebenstreitThesis}. Here
$n(\s{p},t)$ is the number density of quasi-particles of momentum $\s{p}$ at time $t$, fulfilling \cite{BlochEtAl99, HebenstreitThesis}
\begin{align}
  \dot n(\s{p}, t) & = Q(\s{p},t) k(\s{p}, t), \nonumber \\
\label{eq:n_eom}
  \dot k(\s{p}, t) & = Q(\s{p},t) (1- n(\s{p}, t)) - 2 \omega(\s{p},t) \ell(\s{p},t), \\
  \dot \ell(\s{p}, t) & = 2 \omega(\s{p},t) k(\s{p}, t), \nonumber
\end{align}
with
\[
 Q(\s{p}, t) = \frac{e E(t) \eps_\perp}{\omega_\s{\pi}^2}
\]
and the initial conditions $n(\s{p}, t_0) = k(\s{p}, t_0) = \ell(\s{p}, t_0) = 0$ for some time $t_0$ before the potential is switched on.

The first term in the integrand in \eqref{eq:QKE_current} gives rise to the so-called \emph{conduction current}, due to the movement of the quasi-particles already created. The last two terms yield the \emph{polarization current}, due to the creation of quasi-particle pairs. The last term is a counterterm, canceling a divergence in the polarization current. This will be crucial in our discussion of the enhanced quasi-particle numbers at intermediate times. Note that the expression \eqref{eq:QKE_current} for the vacuum current is the correct expression for quasi-particle densities as defined in the QKE framework. For other definitions of quasi-particles, the expression has to be modified (but note the discussion in Section~\ref{sec:Discussion} below). 

As discussed in Appendix~\ref{app:RelationQKEPointSplit}, the equations \eqref{eq:j} and \eqref{eq:QKE_current} are equivalent, the precise relation being given by \eqref{eq:nRelation}.

\begin{figure}
\centering
\includegraphics[width=0.8\textwidth]{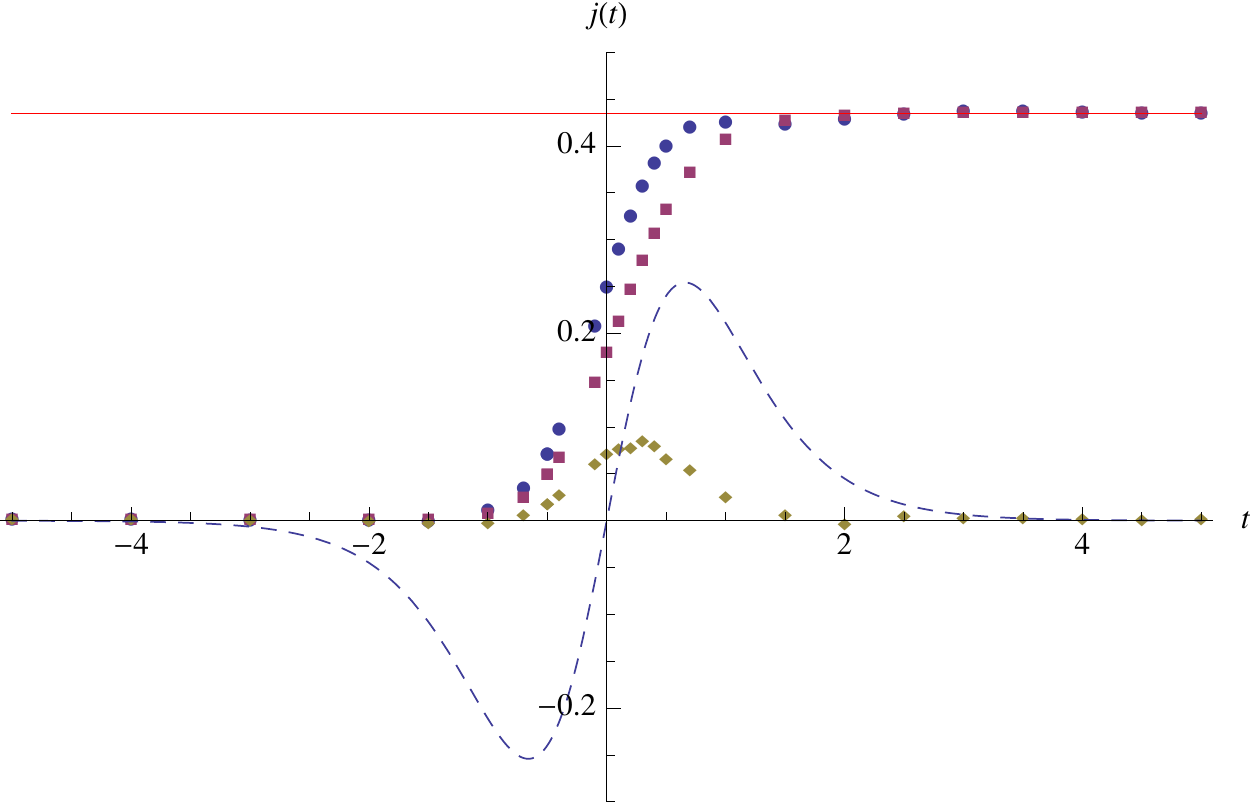}
\caption{The vacuum current density (blue discs) as given by \eqref{eq:j} for the super-critical peak field strength $E_0 = 10 E_c$ and $m=w=1$. The red line indicates the asymptotic value. Also shown are the conduction (violet squares) and polarization (yellow diamonds) current densities, \cf \eqref{eq:QKE_current}. For comparison, also the external current density, divided by 100, is shown (blue dashed curve). All quantities are expressed in natural Compton units.}
\label{fig:Tanh10}
\end{figure}

\begin{figure}
\centering
\includegraphics[width=0.8\textwidth]{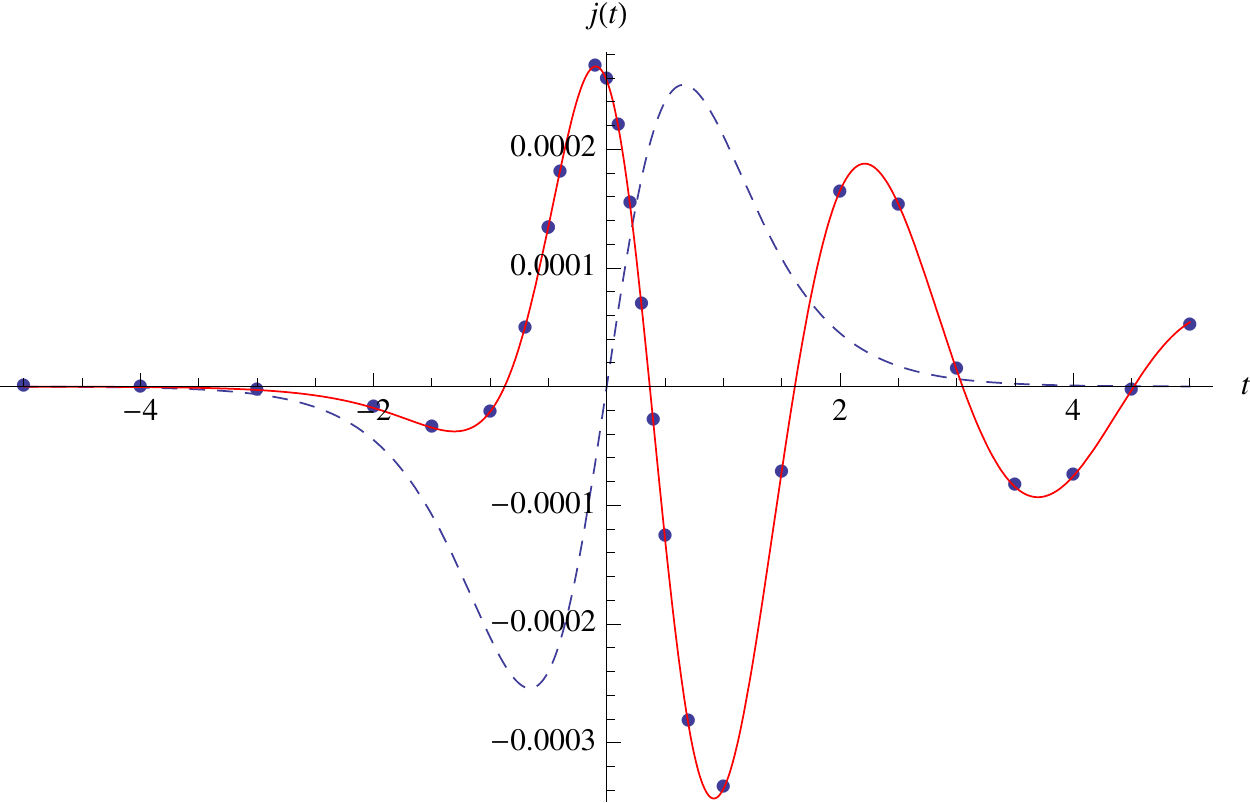}
\caption{The vacuum current density (blue discs) as given by \eqref{eq:j} for the  sub-critical peak field strength $E_0 = 0.1 E_c$ and $m=w=1$. Now the red line is the prediction from Serber's equation \eqref{eq:Serber}. The external current density, divided by 1000, is given by the dashed blue line.}
\label{fig:Tanh01}
\end{figure}

The mode solutions $\psi^{\pm}_{\s{p}, s}$ for the Sauter pulse \eqref{eq:Potential} are explicitly known \cite{NarozhnyiNikishov70}, so the expression \eqref{eq:j} can be evaluated numerically rather straightforwardly.\footnote{This is especially the case in the absence of a hierarchy of scales, i.e., when the ``frequency'' $w$ is of the same order of magnitude as the mass $m$. For simplicity, we restrict to that case. Investigations with longer pulse durations show that the results are qualitatively similar, with the exception of the damping of the oscillating tail in the sub-critical case, \cf Fig.~\ref{fig:Tanh01} and the discussion at the end of Section~\ref{sec:Linearization}.}  
For super-critical peak field strengths $E_0 > E_c$, we find that the vacuum current density at intermediate times increases quasi-monotonically to its asymptotic value, even for very short pulses, 
\cf Fig.~\ref{fig:Tanh10}. Furthermore, the slope of the vacuum current density is largest approximately where the field strength has its maximum, as one might expect. Note that also the asymptotic value is shown, as computed by integrating the pair production probability density $P(\s{p})$ of Narozhnyi and Nikishov \cite{NarozhnyiNikishov70} with the group velocity $v(\s{p})$. The vacuum current density nicely converges to this asymptotic value. Note that, as indicated in the figure, the vacuum current density is two orders of magnitude smaller than the external one. Hence, the neglection of back-reaction effects seems justified, at least on the time scales considered here.\footnote{In \cite{BlochEtAl99} a seemingly different conclusion was obtained. However, the electric charge was there set to $e^2 = 4$ instead of the value $e^2 = 4 \pi/137$ used here.} Also shown in the figure are the conduction and polarization currents as computed by \eqref{eq:QKE_current} using the quasi-particle densities $n(\s{q},t)$ for the Sauter pulse \cite{HebenstreitThesis}. 

For sub-critical peak field strengths $E_0 < E_c$, we find that the vacuum current density at intermediate times can be much larger than at asymptotically late times, \cf Fig.~\ref{fig:Tanh01}, where the asymptotic current density is about three orders of magnitudes smaller than at intermediate times. Furthermore, for very short pulses, it oscillates after the passing of the pulse with a period of about half the Compton time of the electron.
Now the vacuum current as defined in the QKE framework is nearly entirely due to polarization (so that the decomposition into the conduction and polarization current is not shown, the polarization current being indistinguishable from the full result).

\section{Linearization in the external field}
\label{sec:Linearization}

One of the first applications of QED in external potentials was Uehling's calculation of vacuum polarization in static potentials, at linear order in the external potential \cite{Uehling35}. A generalization for the time-dependent case is due to Serber \cite{Serber35}. However, his derivation is not entirely correct, \cf Appendix~\ref{app:Serber}. A corrected version is
\begin{equation}
\label{eq:Serber}
 \langle j^\mu(x) \rangle = \int \ud^4 x' \ K(x-x') \Box \Box J^\mu(x'),
\end{equation}
where $J$ is the external current density and $K$ vanishes outside of the forward light cone and is given by
\begin{equation}
\label{eq:K}
 K(x) = \frac{\alpha}{8 \pi^2 \sqrt{x^2}} \int_0^{\pi/2} \ud \psi \ \cos^4 \psi \int_m^\infty \ud k \ J_1(2 k \sqrt{x^2} / \cos \psi) / k^2
\end{equation}
inside, with $J_1$ being a Bessel function. Interestingly, \eqref{eq:Serber} only depends on the external current, not on the field strength. Heuristically, in time-dependent situations, the vacuum current \eqref{eq:Serber} can be thought of as the current due to a rearrangement of the Uehling vacuum polarization clouds accompanying the charges that are responsible for the external current. In this sense, it is indeed a polarization current, as in the terminology of \cite{BlochEtAl99}, discussed above. However, it may be misleading to attribute it to pair creation, as usually done. An expression similar to \eqref{eq:Serber} was found by Schwinger \cite{SchwingerQED2}, who, however, assumed an external potential vanishing in the past and the future, a condition not fulfilled in the present case.

As shown in Figure~\ref{fig:Tanh01}, the expression \eqref{eq:Serber} describes the time-dependence of the vacuum current in the sub-critical case very well. We note that in the example shown in Figure~\ref{fig:Tanh01} the adiabaticity parameter $m \omega / \betrag{e E}$ is larger than one, indicating that the (asymptotic) pair production probability is dominated by multi-photon processes \cite{BrezinItzykson70}. Nevertheless, the vacuum current density at intermediate times is very well described by the linear approximation. We also note that the asymptotic value of Serber's expression \eqref{eq:Serber} for the vacuum current density vanishes. This follows from the Riemann-Lebesgue lemma and the integrability of the Fourier transform of $K$ at its singularity $p^2= 4 m^2$.

\begin{figure}
\centering
\includegraphics[width=0.8\textwidth]{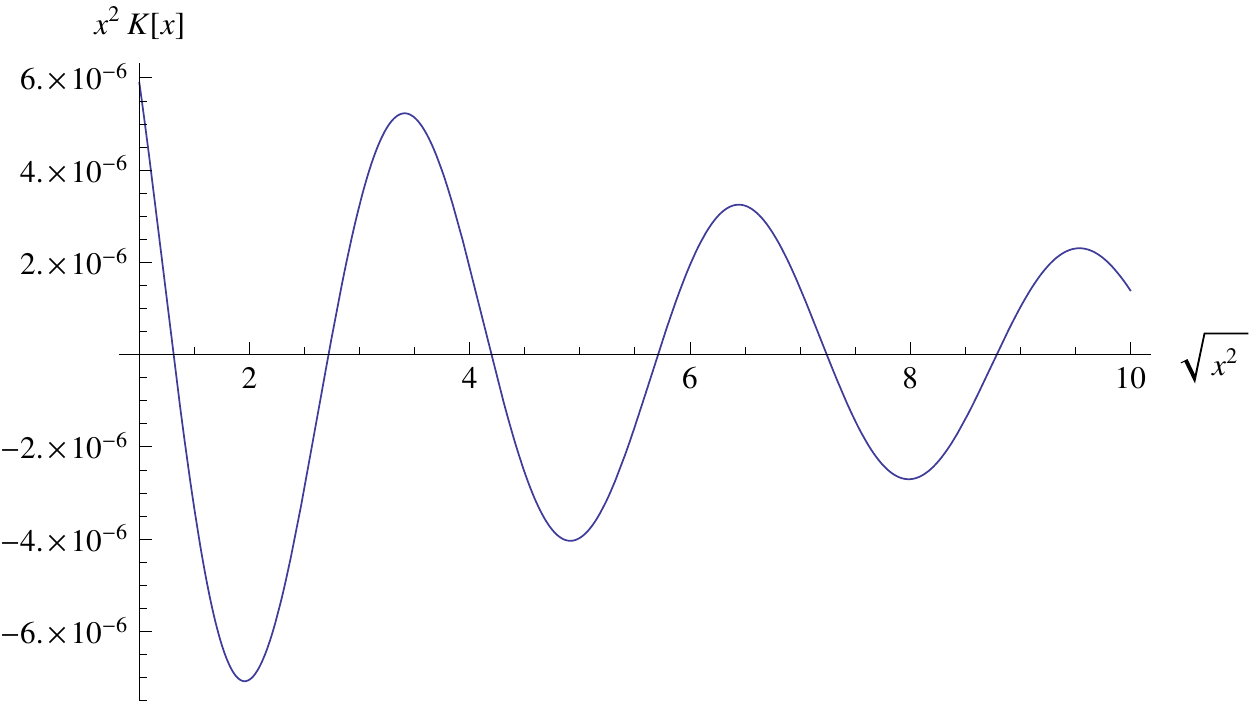}
\caption{Serber's kernel $K(x)$, \cf \eqref{eq:K}, multiplied by $x^2$. Distances are measured natural Compton units.} 
\label{fig:K}
\end{figure}

The oscillatory behavior of the current shown in Fig.~\ref{fig:Tanh01} is a manifestation of the oscillatory nature of $K$, \cf Fig.~\ref{fig:K}. Clearly, for external currents that vary on time-scales much larger than the Compton time of the electron, these oscillations are strongly damped. Following Schwinger \cite{SchwingerQED2}, one may then write the vacuum current density as
\begin{equation*}
 \langle j^\mu \rangle = \sum_{n=1}^\infty c_n m^{-2n} \Box^n J^\mu,
\end{equation*}
for certain constants $c_n$ (which can also easily be obtained by expanding the function $\chi(k)$ of Serber \cite{Serber35} around $k=0$). 

\section{The enhanced quasi-particle densities at intermediate times}
\label{sec:Discussion}

Let us now turn to the discussion of the enhanced quasi-particle densities described in \cite{BlaschkeEtAl06}.
There, a slowly varying sub-critical electric field $E = E_0 \sin w t$ is considered, which is turned on and off at some finite times. At first sight, one might think that the explanation lies in the fact that for sub-critical electric fields the vacuum current density at intermediate times is dominated by the linearization \eqref{eq:Serber}, which is known to vanish at asymptotic times. A first indication that this reasoning is flawed is the fact that one may well have non-vanishing particle densities at asymptotic times in spite of a vanishing vacuum current. As we will see, a satisfactory explanation of the enhanced quasi-particle densities at intermediate times has to take the logarithmic divergence of the current into account.

\begin{figure}
\centering
\includegraphics[width=0.8\textwidth]{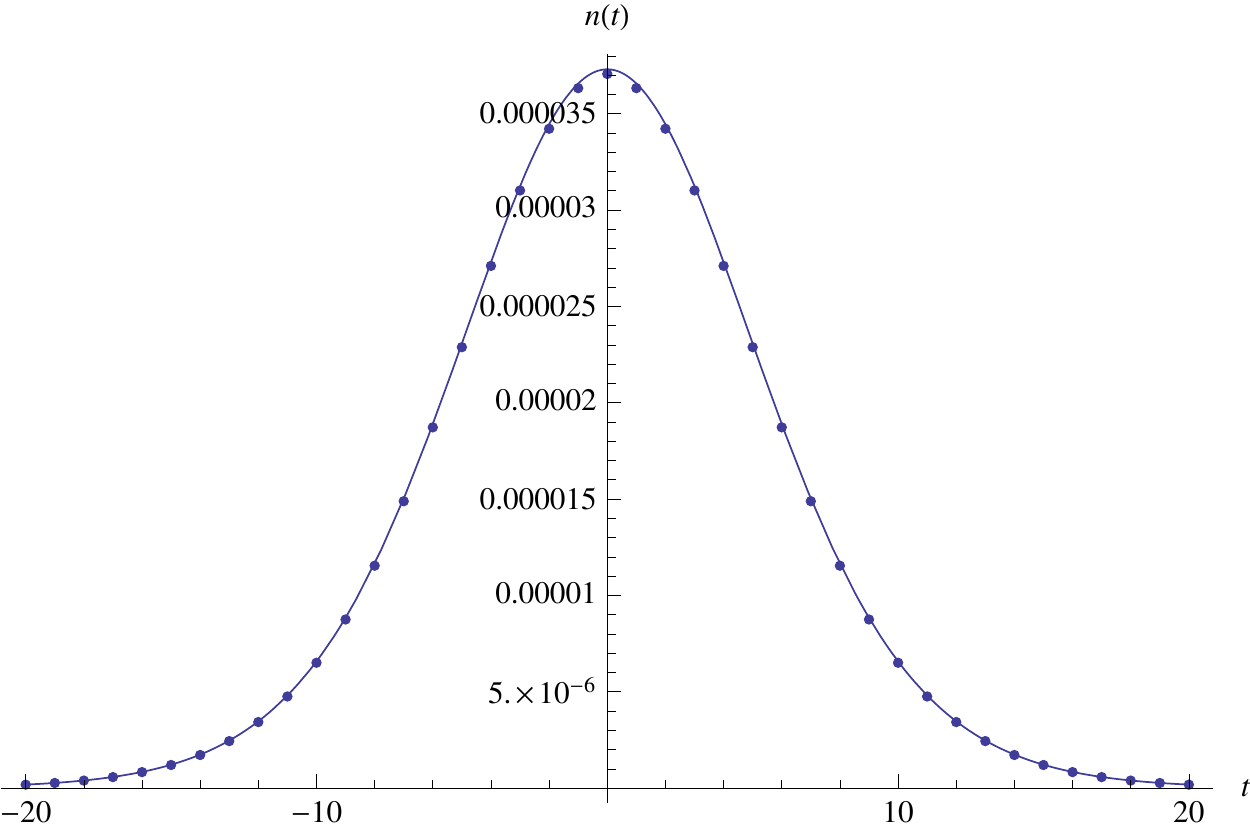}
\caption{The quasi-particle number density as defined in the framework of the quantum kinetic equation (blue discs), compared to the expectation \eqref{eq:n}, for the Sauter pulse \eqref{eq:Potential} with $E_0 = 0.1 E_c$, $w = 0.1$, $m=1$.}
\label{fig:ParticleNumber}
\end{figure}

We recall that the third term in \eqref{eq:QKE_current} is a counterterm which is necessary to cancel a logarithmic divergence in the polarization current, i.e., the integral over the second term in \eqref{eq:QKE_current}. It follows that for large momentum $\s{p}$ the rate of the change of the quasi-particle density is given by\footnote{Actually, in \cite{BlochEtAl99}, the counterterm in \eqref{eq:QKE_current} is derived from this asymptotic form of $\dot n$.}
\[
 \dot n(\s{p}, t) \sim e^2 E(t) \dot E(t) \frac{\eps_\perp^2}{4 \omega(\s{p}, t)^6}.
\]
For slowly varying sub-critical external fields, one expects that this yields the main contribution to the rate of change of the total quasi-particle density. 
Performing the momentum integral over the fraction on the \rhs and integrating in time, this implies
\begin{equation}
\label{eq:n}
 n(t) \sim \tfrac{3}{256 \pi} e^2 E(t)^2 m^{-1}.
\end{equation}
Up to a factor $\frac{3 \pi^2}{32} \simeq 0.93$, which seems to have been interpreted as $1$, this agrees with the heuristic description of the numerical results found in \cite{BlaschkeEtAl06} for a sinusoidal time-dependence of the background electric field.
The expectation \eqref{eq:n} can also be verified for slow sub-critical Sauter pulses \eqref{eq:Potential}, using the quasi-particle number density from \cite{HebenstreitThesis}, \cf Figure~\ref{fig:ParticleNumber}. In this sense, the enhanced quasi-particle densities reported in \cite{BlaschkeEtAl06} are primarily due to the fact that the dynamical variables chosen in the QKE framework are such that a supplementary counterterm is necessary in the expression for the vacuum current density, i.e., that $\dot n$ is much larger than it would need to be in order to yield the correct vacuum current density.

A strategy to circumvent the problem of greatly enhanced particle number densities would thus be to find dynamical variables for which no further renormalization is necessary in the computation of the current density. One possibility would be to consider better adapted modes for the definition of quasi-particle densities, perhaps similar to those discussed in \cite{DabrowskiDunne14} for the charged scalar field. However, it seems unlikely that this is possible: we know that the $\beta$ function of QED does not vanish, so there must be a logarithmic divergence in the current, proportional to the external current $J = - \dot E$ (a change $j \to j + \lambda J$ corresponds to a charge renormalization). Hence, for any definition of quasi-particle densities, one expects a logarithmic divergence proportional to $\dot E$ in the corresponding expression for the vacuum current.

Instead, one might proceed as follows: The essence of Dirac's procedure is to subtract from the two-point function $\langle \psi(x) \bar \psi(y) \rangle$ its singular part $H(x, y)$, the Hadamard parametrix. The result $R(x,y)$ is smooth, so that the limit of coinciding points may be performed, for example in the computation of the expectation value of the current density or the stress-energy tensor. The idea would be to solve the differential equation for the smooth remainder $R$,\footnote{It may be advantageous to restrict to coinciding times and to include a compensating parallel transport in the definition of $R$, similarly to the Wigner function framework \cite{BialynickiBirulaEtAl91}.} i.e.,
\begin{align}
\label{eq:R1}
 D R(x,x') & = D H(x,x'), &
 {D'}^t R(x,x') & = {D'}^t H(x,x'),
\end{align}
where ${D'}^t$ is the transpose of $D = i \nabslash-m$ (acting on a co-spinor), applied on the primed variable. Here we used that the two-point function is a bi-solution to the Dirac equation. In contrast, the parametrix is not a bi-solution, but there are smooth remainders which appear as source terms on the right hand side. These remainders are given \cite{ChiralFermions} as a series in the Hadamard coefficients $V_k(x,x')$ (the Lorentzian analogues of the heat kernel coefficients) for the wave operator
\[
 P = \nabla^\mu \nabla_\mu + \tfrac{i e}{2} \gamma^\mu \gamma^\nu F_{\mu \nu} + m^2.
\]
Instead of a linear differential equation for the Bogoliubov coefficients, one thus has to solve an inhomogeneous one for the remainder $R$. The renormalization ambiguity of the current is now hidden in the initial condition for $R$ (choosing $R = 0$ in the past amounts to Heisenberg's renormalization condition \eqref{eq:Lambda}). The vacuum current density $\langle j^\mu(x) \rangle$ is simply given by $\tr \gamma^\mu R(x, x)$, i.e., no further renormalization is necessary.  For the solution $R$ one would thus expect that the magnitude at intermediate times is of the same order as asymptotically. Once the background field is switched off, and hence $H$ equals the vacuum two-point function, one can read off the particle content directly from the Fourier transform of $R$. 

A nice feature of this scheme is that, due to the initial condition $R=0$, all pair creation effects are due to the inhomogeneous part on the \rhs of \eqref{eq:R1}. Hence, it may be possible to make statements about pair creation without actually solving the differential equations \eqref{eq:R1}.\footnote{Note, however, that the Hadamard coefficients $V_k$ that appear on the \rhs of \eqref{eq:R1} are recursively defined by an inhomogeneous ODE, \cf \cite{Current} for details.} A further advantage of the framework is that it is in principle applicable to generic external potentials, in contrast to the QKE, which requires a homogeneous, purely electric field. 

On the other hand, there are obvious difficulties: the construction of $H$ involves a series
\[
 \sum_k V_k(x,x') (x-x')^{2 (k-1)},
\]
which is only guaranteed to converge for analytic external fields. In practice, one would probably truncate the series, so that $R$ is not smooth (but $C^m$ for arbitrarily large $m$). This potentially leads to problems at large distances, so it might be necessary to include an infrared cut-off in the definition of $H$.
A further difficulty lies in the fact that the Hadamard coefficients $V_k$ are not known explicitly to all orders (however, for very symmetric external fields it could be possible to fully determine them). In view of the potential benefits, we consider it worthwhile to try to overcome these difficulties.

We remark that the strategy proposed above is quite similar to the modified equal time Wigner function formalism\footnote{Furthermore, in the context of the back-reaction problem of the semi-classical Einstein equation, a similar approach was taken in \cite{EltznerGottschalk} for the evaluation of the stress-energy tensor.} used in \cite[Sec.~5.2]{HebenstreitThesis}. There, one subtracts from the two-point function not the Hadamard parametrix, but the vacuum two-point function, multiplied by a parallel transport. This may be appropriate in the 1+1 dimensional setting considered there, as the remainder is continuous (though not continuously differentiable) for equal times. However, it easily follows from the results in \cite{Current}, that in 3+1 dimensions the remainder would not be continuous. In particular, the logarithmic divergence that is responsible for the charge renormalization ambiguity is not canceled, so one expects the problems with greatly enhanced values at intermediate times to persist.

\section{Conclusion}

To summarize, we have introduced a method to compute the expectation value of the current density \eqref{eq:CurrentDensity} in QED in external time-dependent potentials at intermediate times, based on a point-splitting renormalization. We showed that it is equivalent to the expression for the current density derived in the QKE framework \cite{BlochEtAl99}. We also corrected an expression due to Serber for the linearization of the vacuum current density in the external potential. Our non-perturbative results for the Sauter pulse \eqref{eq:Potential} are compatible both with the corrected version \eqref{eq:Serber} of Serber's general expression in the case of a sub-critical peak field strength and with the asymptotic results of Narozhnyi and Nikishov \cite{NarozhnyiNikishov70}. 

We proposed a natural explanation of the enhanced quasi-particle densities at intermediate times for slowly varying sub-critical homogeneous electric fields \cite{BlaschkeEtAl06}. The argument used the fact that the current is completely fixed up to the charge renormalization ambiguity. We also sketched a framework in which no difference of scales of the dynamical quantity at intermediate and asymptotic times is to be expected, and which is in principle applicable to generic external potentials. The development of this scheme is a topic for future work.

\subsection*{Acknowledgements}
I would like to thank Sergey Gavrilov, Holger Gies, Jan Schlemmer and Ralf Sch\"utzhold for helpful discussions or hints to the literature. Parts of this work were done at the Faculty for Physics of the University of Vienna, supported by the Austrian Science Fund (FWF) under the contract P24713.

\appendix

\section{Comparison of point-splitting and QKE}
\label{app:RelationQKEPointSplit}

To fix conventions, we choose signature $(+, -, -, -)$ and the standard Dirac matrices. The covariant derivative is defined as
\[
 \nabla_\mu = \del_\mu + i e A_\mu.
\]
We construct a Klein-Gordon type operator
\[
 P = (i \nabslash - m) (- i \nabslash - m) = \nabla_\mu \nabla^\mu + i \tfrac{e}{2} \gamma^\mu \gamma^\nu F_{\mu \nu} + m^2 
\]
As only one independent component of $F_{\mu \nu}$ is non-vanishing, namely $F_{0 3}$, we choose two linearly independent eigenvectors
\begin{align*}
 \Gamma^1 &= \begin{pmatrix} 0 \\ 1 \\ 0 \\ -1 \end{pmatrix}, &
 \Gamma^2 &= \begin{pmatrix} 1 \\ 0 \\ 1 \\ 0 \end{pmatrix}
\end{align*}
of $\gamma^0 \gamma^3$ with eigenvalue $1$. On these, the operator $P$ acts as the scalar operator
\[
 P = \nabla_\mu \nabla^\mu + i e F_{03} + m^2.
\]
We assume $A_\mu = \delta_{3 \mu} A(x^0)$ and look for solutions $g_\s{p}$ of the form $g_{\s{p}}(x) = f_\s{p}(x^0) e^{i \s{p} \s{x}}$. Hence, $f_\s{p}$ fulfills
\begin{equation}
\label{eq:f_eom}
 \ddot f_\s{p} + \left( \s{\pi}^2 + m^2 + i e E \right) f_\s{p} = 0,
\end{equation}
with $\pi$ as in \eqref{eq:def_pi}.
Assume that $A(t)$ is switched on at some finite time and let $f_\s{p}(x^0) = e^{-i \omega_\s{p} x^0}$ in the past. Properly normalized spinorial solutions to the Dirac equation are then given by
\[
 \psi^+_{\s{p},s} = \frac{1}{2 (2\pi)^{3/2}} \frac{1}{\sqrt{\omega_{\s{p}} (\omega_{\s{p}} - p_3)}} (- i \nabslash - m) g_{\s{p}} \Gamma^s.
\]
Using that
\[
 \sum_s \Gamma^s \bar \Gamma^s = \gamma^0 - \gamma^3,
\]
one easily obtains
\begin{equation}
\label{eq:f_gamma3_f}
 \sum_s \overline{f^+_{\s{p}, s}(x^0)} \gamma^3 f^+_{\s{p}, s}(x^0) = \frac{( \eps_\perp^2 - \pi_3^2 ) \betrag{f_\s{p}}^2 - \betrag{\dot f_\s{p}}^2 - 2 \pi_3 \Im( \bar f_\s{p} \dot f_\s{p}) }{(2\pi)^3 \omega_\s{p} (\omega_\s{p} - p_3)}.
\end{equation}
On the other hand, for the vacuum modes evaluated at the shifted momentum, we have
\begin{equation}
\label{eq:f0_gamma3_f0}
 \sum_s \overline{f^{0,+}_{\s{\pi}, s}(x^0)} \gamma^3 f^{0,+}_{\s{\pi}, s}(x^0) = \frac{2 \pi_3}{(2\pi)^3 \omega_\s{\pi}}.
\end{equation}
Furthermore, it follows from \eqref{eq:f_eom} that
\begin{align}
 \del_t \betrag{f_\s{p}}^2 & = 2 \Re ( \bar f_\s{p} \dot f_\s{p}), \nonumber \\
 \label{eq:f_p_eom}
 \del_t (\bar f_\s{p} \dot f_\s{p}) & = \betrag{\dot f_\s{p}}^2 - (\omega_{\s{\pi}}^2 + i e E) \betrag{f_\s{p}}^2, \\
 \del_t \betrag{\dot f_\s{p}}^2 & = - 2 \omega_\s{\pi}^2 \Re ( \bar f_\s{p} \dot f_\s{p}) - 2 e E \Im ( \bar f_\s{p} \dot f_\s{p}). \nonumber
\end{align}
Comparison of \eqref{eq:QKE_current} with \eqref{eq:f_gamma3_f}, \eqref{eq:f0_gamma3_f0} suggest that
\begin{equation}
\label{eq:QKE_f_p}
 \frac{\pi_3}{\omega_\s{\pi}} n + \frac{\omega_\s{\pi}}{e E} \dot n = \frac{\pi_3}{\omega_\s{\pi}} - \frac{( \eps_\perp^2 - \pi_3^2 ) \betrag{f_\s{p}}^2 - \betrag{\dot f_\s{p}}^2 - 2 \pi_3 \Im( \bar f_\s{p} \dot f_\s{p}) }{2 \omega_\s{p} (\omega_\s{p} - p_3)}.
\end{equation}
Indeed, using \eqref{eq:f_p_eom}, one can show that with
\begin{equation}
\label{eq:nRelation}
 n = 1 + \frac{\pi_3 \left( \omega_\s{\pi} \betrag{f_\s{p}}^2 + \frac{1}{\omega_\s{\pi}} \betrag{\dot f_\s{p}}^2 \right) + 2 \omega_\s{\pi} \Im(\bar f_\s{p} \dot f_\s{p}) }{2 \omega_\s{p} (\omega_\s{p} - p_3)}
\end{equation}
one fulfills not only \eqref{eq:QKE_f_p} but also the system of equations \eqref{eq:n_eom}, with the correct initial conditions. Furthermore, it is straightforward to verify that
\begin{equation}
\label{eq:CountertermRelation}
 \frac{1}{(2 \pi)^3} \int \ud^3 p \frac{\eps_\perp^2}{2 \omega_\s{p}^5} e^{i p_3 z} = - \frac{1}{12 \pi^2} \left( \log \frac{z^2 m^2 e^\gamma}{4} - 1 \right) + \order(z). 
\end{equation}
This proves the equivalence of \eqref{eq:j} and \eqref{eq:QKE_current}.

\section{The correction of Serber's result}
\label{app:Serber}

Let us recall the arguments of Serber \cite{Serber35} in his derivation of the vacuum current at linear order in the external field. He finds that, in units where $m=1$,
\[
 j_\mu(x) = \frac{\alpha}{32 \pi^5} \int_0^{\pi/2} \cos^3 \psi \ud \psi \int \Lambda(x-x', \psi) \Box \Box J_\mu(x') \ud^4 x',
\]
where\footnote{To obtain the retarded response, one has to replace $k^2$ in the $\log$ and the denominator by $k^2 + i \eps k_0$. Serber does not do this at this stage and obtains the mean of the retarded and advanced response, only selecting the retarded response at the very end (and multiplying it by $2$). This explains the missing factor of $2$ in \eqref{eq:App_Lambda_2} with respect to \eqref{eq:K}.}
\begin{equation}
\label{eq:App_Lambda_1}
 \Lambda(x, \psi) = \int e^{-i k x} \log(1 - \tfrac{1}{2} k^2 \cos \psi) k^{-4} \ud^4 k,
\end{equation}
\cf equations $\{ 13 \}$ and $\{ 14 \}$ (equation numbers in curly brackets refer to \cite{Serber35}). It is then shown that $\Lambda$ vanishes outside of the light cone and is given by
\begin{equation}
\label{eq:App_Lambda_2}
 \Lambda(x,\psi) = \frac{2 \pi^3 \cos \psi}{\sqrt{x^2}} \int_1^\infty J_1(2 k \sqrt{x^2} / \cos \psi) k^{-2} \ud k  
\end{equation}
inside, \cf equation $\{ 17 \}$. To determine the value on the boundary of the forward light cone, assumed to be of the form $f(r, \psi) [ \delta(t - r) + \delta(t+r)]$, Serber proceeds as follows: He first argues that the integral of \eqref{eq:App_Lambda_2} over $x^0$, for any fixed spatial $x^i$, vanishes. He then computes the same integral for \eqref{eq:App_Lambda_1}, finds a non-vanishing result, and concludes that it must be due to a non-vanishing $f$, i.e., a contribution localized on the boundary of the light cone. However, both these integrals are not computed correctly. Regarding the vanishing of the time integral over \eqref{eq:App_Lambda_2}, Serber first introduces the variable $s=\sqrt{t^2-r^2}$ and correctly notes that \cite[Sect.~13.6]{Watson44}
\[
 \int_0^\infty J_1(2 k s / \cos \psi) / \sqrt{s^2+r^2} \ud s = I_{\frac{1}{2}}(k / \cos \psi) K_{\frac{1}{2}}(k / \cos \psi),
\]
but then argues that this vanishes, due to $K_{\frac{1}{2}}(z) = 0$. However, it is well-known \cite[Sect.~10.2]{AbramowitzStegun} that $K_{\frac{1}{2}}(z) = \sqrt{\pi/(2z)} e^{-z}$. Using computer algebra, the time integral of \eqref{eq:App_Lambda_2} can be computed explicitly, yielding
\begin{multline*}
 I_{\eqref{eq:App_Lambda_2}}(r, \psi) = \frac{4 \pi^3}{r} \Big[ \tfrac{1}{4} \cos^2 \psi + ( \tfrac{1}{2} r \cos \psi - \tfrac{1}{4} \cos^2 \psi) e^{-2r/\cos \psi} \\ + r^2 \Ei(-2 r / \cos \psi) \Big],
\end{multline*}
where $\Ei$ denotes the exponential integral. For the time integral of \eqref{eq:App_Lambda_1}, Serber claims that
\begin{align*}
 I_{\eqref{eq:App_Lambda_1}}(r, \psi) & = \frac{8 \pi^2}{r} \int_0^\infty \sin(kr) \log (1 + \tfrac{1}{4} k^2 \cos^2 \psi) / k^3 \ud k \\
 & = - \frac{2 \pi^3 \cos^2 \psi}{r} \int_1^\infty e^{- 2 k r / \cos \psi} / k^3 \ud k,
\end{align*}
the step from the first to the second line being the result of a deformation of the integration path to the imaginary axis. However, this is not straightforward, as the closing of the path at infinity diverges, due to the $\sin$. Numerical integration shows that integrals in the first and the second line do not coincide, so the path deformation seems to have been performed incorrectly. Performing the integral in the first line by computer algebra, one finds that $I_{\eqref{eq:App_Lambda_1}}(r, \psi) = I_{\eqref{eq:App_Lambda_2}}(r, \psi)$. Hence, no supplementary term localized on the boundary of the light cone is present.


\end{document}